\documentclass[twocolumn,showpacs,preprintnumbers,amsmath,amssymb]{revtex4}

\usepackage{graphicx}
\usepackage{epsfig}
\usepackage{bm}% bold math

\usepackage{amssymb}

\begin{document}

\title{ Volume fluctuations and geometrical constraints in granular packs  }

\author{ Tomaso Aste }
\affiliation{ Department of Applied Mathematics, RSPhysSE, The Australian National University, 0200 Australia.}

%\author{}
\date{ \today }% It is always \today, today,

\begin{abstract}
% Text of abstract
Structural organization and correlations are studied in very large packings of equally sized acrylic spheres, reconstructed in three-dimensions by means of X-ray computed tomography.
A novel technique, devised to analyze correlations among more than two spheres, shows that the structural organization can be conveniently studied in terms of a space-filling packing of irregular tetrahedra.
The study of the volume distribution of such tetrahedra reveals an exponential decay in the region of large volumes; a behavior that is in very good quantitative agreement with theoretical prediction.
I argue that the system's structure can be described as constituted of two phases: 
1) an `unconstrained'  phase which freely shares the volume; 
2) a `constrained'  phase which assumes configurations accordingly with the geometrical constraints imposed by the condition of non-overlapping between spheres and mechanical stability.
The granular system exploits heterogeneity maximizing freedom and entropy while constraining mechanical stability.
\end{abstract}


% main text

\maketitle

The study of how space is shared among the particles in a granular pack is essential for understanding both the static properties of these structures and the dynamical mechanisms which generate them. 
When equal spheres are packed in a container they can arrange in a way to minimize potential  (gravitational) energy by maximizing the packing fraction. 
The pursuit of maximum compaction is common to several physical systems and, at atomic level, it is a feature associated with metallic bounding. 
From a purely geometrical perspective, it is known that the largest attainable packing fraction in a system of equal spheres is  $\rho = \pi/\sqrt{18} \sim 0.74$    \cite{Conway97,ppp}; which corresponds to a disposition of parallel hexagonal layers of spheres  in stacks (forming the so-called Barlow packings). 
Conversely, it is observed empirically that when balls are poured in a container they spontaneously arrange in a disorderly fashion occupying a fraction of the total volume between 0.555 and 0.64. 
The study of these disordered structures is very challenging and the available investigation tools appear to be inadequate to capture their essential features.  
Indeed, a complete description of the structure of a disordered system requires a very large amount of information about the coordinates, orientations, shapes and connectivities of all the elements. 
It is however clear that not all this information is necessary to determine the   properties of these systems.
On the contrary, there exist  several states with different microscopic realizations which share the same macroscopic properties. 
One of the challenges of the research in this field is to find a simple measure which characterizes the state of the system giving information about the packing structure and its properties \cite{Barrat00,Makse02,Fierro02,Dean03,DAnna03,Schroder05}.

It has been argued by Edwards \cite{Edwards89,Edwards94}  that granular systems can be described in terms of a Gibbs-like equilibrium thermodynamical approach where the conservation of energy  is replaced with a constraint on the total volume $V$.
In this approach,  the state of the system can be described by a state variable, the `compactivity', which plays the role of temperature.
Such compactivity is associated with the volume $V$ and therefore to the packing fraction $\rho$.
A key issue is to establish the appropriate elementary volumes which are capable to fully describe the system as a whole  \cite{Edwards04}.
In ref.\cite{Edwards04} such volumes are the polyhedra constructed from the first coordination shell. 
The theory predicts that the distribution of such elementary volumes must follow the exponential law $\exp[-v/(\lambda X)]$ with $X$ the `compactivity' and $\lambda$ a constant analogous to the Boltzmann one.

\begin{figure}
\begin{center}
\resizebox{0.4\textwidth}{!}{%
\includegraphics{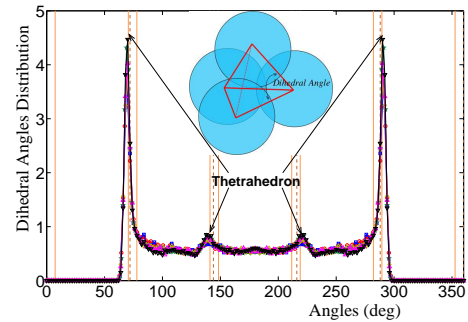} 
}
\end{center}
\caption{ Dihedral angle distribution (x-axis: angular degrees; y-axis: renormalized frequencies).
The vertical lines indicate the angles  $\theta = n \arccos(1/3)$ (and $360 -\theta$) with $n = 1,2,3,4,5$ (tetrahedral packings).
The dashed lines are at the angles $\theta = n 360/5$ ($n = 1, 2, 3, 4$).}
\label{f.dihe}
\end{figure}

In this paper, I show that  the `state' of a disordered  sphere pack can be described in term of a simple parameter  which depends on the packing fraction and on a simple topological property.
This is shown by \emph{First} searching for the local structural motifs which make the `building blocks' of such systems.
\emph{Second}, by exploring the allowed local fluctuation of the volumes of such  building blocks, and predicting the volume distribution. 
\emph{Third}, by comparing the theoretical predictions with the experimental results for equal spheres packings.

The experimental data reported in this paper  are based on the analysis of the largest empirical dataset on disordered structures presently available in the literature \cite{AstePRE05}. 
Such a dataset  is constructed from the study, by means of X-ray Computed Tomography, of large samples of disorderly packed monosized spheres.
This database records the positions of more than 385~000 sphere centers from 6 samples of monosized acrylic beads prepared in a cylindrical container. 
The precision on the coordinates is better than 0.1~\% of the sphere-diameters and the sphere polydispersity is within 2~\%.
In this paper we refer to these samples with labels A, B, C, D, E and F; their packing fractions and sizes are reported in Table. \ref{t.1}. 
The investigations reported in this paper are performed over an internal region ($\mathbf G$) 4 sphere-diameters away from the sample boundaries. 
(Spheres outside $\mathbf G$ are considered when computing the neighboring environment of spheres in $\mathbf G$ ).

\begin{table}
\begin{tabular}[c]{lllll}
\hline
 & packing fraction  & $\;\;\;N$ & $\;\;N_G$  & $\;\left< f \right>$  \\
\hline
{\bf A} & $0.586  \pm 0.005$ & 102897  & 54719 &  14.6\\
{\bf B } & $0.596 \pm 0.006$ & 34016    & 15013 &  14.6\\
{\bf C} & $0.619\pm 0.005$ & 142919  & 91984   & 14.4\\
{\bf D} & $0.626 \pm 0.008$ & 35511    & 15725  & 14.4\\
{\bf E} & $0.630 \pm 0.01$   & 35881    & 15852   & 14.4\\
{\bf F}  & $0.640 \pm 0.005$ & 36461    & 16247  & 14.3\\
\hline
\end{tabular} 
\caption{ 
\label{t.1} Sample density and their intervals of variations ($\pm$) within each sample; 
 total number of spheres  ($N$); 
 number of spheres in the central region ($N_G$); 
 average number of incident Delaunay neighbours $\left< f \right>$. 
}
\end{table}

The search for the elementary building blocks in which the system can be conveniently subdivided and analyzed is performed by introducing a new technique to investigate the structural correlations among the packed spheres.
This analysis is based on two important definitions: \emph{bounded}  spheres and \emph{common neighbor} \cite{Clarke93}.
In particular, two spheres are `bounded' if they stay within a given threshold radial distance $\tilde r$.
Whereas, a `common neighbor' of two bounded sphere is a third sphere which is also bounded to both the two spheres. 
It can be calculated that the maximum number of common neighbors which can be placed around two bounded spheres is equal to 5 for any threshold distance smaller than $ \tilde r  \le \sqrt{5/4}d\simeq 1.118 d$, where $d$ is the spheres' diameter.
Fixed a threshold distance, the numbers of configurations with $n$ common neighbors is a very sensitive measure of local correlations.
For instance, when the threshold distance is $1.1 d$, the fraction of configurations with 4 common neighbors increase sensibly during compaction varying from 17~\% at $\rho = 0.586$ (A) to 31~\% at $\rho = 0.64$ (F).
Similarly, the configurations with 5 common neighbors grow from less than 3~\% to above 8~\% when packing fraction varies between 0.586 to 0.640 (A to F) .

The fact that the number of common neighbors is so sensitive to the packing properties  suggests that
 the study of the local organization around couples of bounded spheres could be the key to  establish which kind of configurations are present in these systems.
To this end, in this paper, I introduce a novel technique to reveal how common neighbors are distributed.
This analysis consists in the study of the dihedral angles between common neighbors around a given couple of bounded spheres. 
These angles are calculated by first constructing the triangle between a couple of  bounded spheres and one common neighbor and then by measuring the dihedral angles between such a triangle and all the other triangles formed with the other neighbors common to the couple of spheres. 
The resulting distribution of angles is shown in Fig.\ref{f.dihe}  (for  threshold distance $\tilde r = 1.1d$).
Such distribution is symmetric in $\theta$ and $360-\theta$ and has two large peaks at 
$\theta = \arccos(1/3) =70.5...$  and $360-\arccos(1/3)=289.4...$.
These values coincide with the dihedral angles in a regular tetrahedron. 
Other two (smaller) peaks are also visible at $\theta=2 \arccos(1/3)=141.0...$ and $\theta = 218.9...$.
They  also correspond to configurations made of two touching regular tetrahera. 
These peaks clearly indicate that the common neighbors tend to gather together forming tetrahedral packings. 
It is worth noting that the essential features of this distribution, and in particular the position of the largest peaks, are little sensitive to the choice of the threshold.
Indeed, the same kind of distributions are obtained for different values of the threshold distance in a range between $1.0d$ and $1.11d$.
The subset of configurations with dihedral angles in the interval within $\arccos(1/3)$  $\pm 1$ degree (and within $360-\arccos(1/3)\pm1$) has been studied in great detail.
It results that -indeed- they are originated by tetrahedral configurations.
In particular, these configurations are very regular tetrahedra with edge-lengths between $0.99d$ and $1.01d$ and volumes which take values within the limits $0.11d^3$ and $0.13d^3$  in 99\% of the configurations (a regular tetrahedron with edge-lengths equal to $d$ has volume $v^*=\sqrt{2}/12d^3 \simeq 0.118 d^3$).

\begin{figure}
\begin{center}
\resizebox{.4\textwidth}{!}{%
\includegraphics{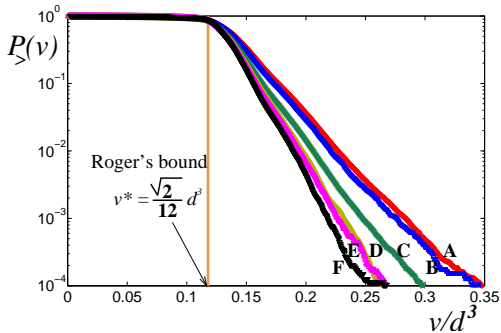}
}
\end{center}
\caption{ 
The inverse normalized cumulants of the distribution of Delaunay volumes in $\mathbf G$ decrease linearly in semilogatithmic scale: $p(v) \sim \exp(-\beta v )$, at large volumes  (best-fits: $\beta d^3 = 43.9; 45.4; 55.2; 64.6; 66.8; 72.9$).
}
\label{f.DeVol}
\end{figure}

The dihedral angles distribution reveals that these amorphous structures can be conveniently viewed as the result of a packing of tetrahedra. 
There is a natural way to subdivide a structure into a system of tetrahedra.
This  is the \emph{Delaunay decomposition} which constructs a system of  minimal tetrahedra with vertices on the centers of neighboring spheres, and chosen in such a way that no other spheres in the pack have centers within the circumsphere of each Delaunay tetrahedron. 
One of the advantages of such decomposition is that it does not require the introduction of any threshold.
The Delaunay decomposition  uniquely associates the packing of $N$ particles with a space-filling systems of a number of  tetrahedra equal to
\begin{equation}
T = \left(\frac{\left< f \right>}{2} -1\right)N \;\;;
\label{T}
\end{equation}
where $\left< f \right>$ is the average number of tetrahedra incident on each particle.
In general,  $\left< f \right>$ takes values in the narrow range between $14 \le \left< f \right> \le 2+48\pi^2/35 \simeq 15.53$.
With the lower limit corresponding to close packed configurations and the upper limit associated with a `granular gas' of randomly positioned particles \cite{Meijering53}.
In mechanically-stable equal-spheres packings, under gravity, this interval of variation reduces further  with typical values around 14.5.
In Table \ref{t.1}, the values of  $\left< f \right>$ for the 6 samples A-F are reported.

Once established that the elementary building blocks which fill the space are tetrahedra, the further step is to explore how these tetrahedra are arranged in space. 
Indeed, some local configurations are closer and others are looser and the whole packing is made by gluing together these tetrahedra in a disordered way which is compatible with the following three conditions: 
\begin{itemize}
\item[1)] mechanical stability; 
\item[2)] geometrical constraints; 
\item[3)] space filling. 
\end{itemize}
Let consider each of these conditions separately.

\emph{Mechanical Stability} is ensured by the network of contact between spheres.
Indeed, in order to equilibrate the number of degree of freedom with the number of constraints, a mechanically stable packing  must satisfy topological conditions on the number of spheres in contact.
In terms of Delaunay decomposition such  topological conditions imply that the average number of Delaunay neighbors must stay in a narrow range around  $\left< f \right> \simeq 14.5$.

The \emph{geometrical constraints} are enforced by the condition of non-overlapping.
Equal spheres pack locally in the closest way when disposed all in touch with each other with centers on  the vertices of a regular tetrahedron.
Such a tetrahedron has volume $v^* = \sqrt{2}/12\;d^3$.
Therefore, the geometrical constraints fix the lower bound for the volume attainable by a Delaunay tetrahedron to $v^*$ (Roger's bound \cite{ppp}). 

The constraint of \emph{space-filling} implies that the sum over all the local volumes of the Delaunay tetrahedra is equal to the total volume ($\sum_i v_i = V$).

If one considers the Delaunay decomposition as an ensemble of $T$ independent cells with volumes $v_i$ that freely exchange volume among each other under the three constraints mentioned above, then the partition function of such a system can be calculated exactly:
$Z = (V-Tv^*)^T / T!$;
% \begin{equation}
%Z = \frac{(V-Tv^*)^T}{T!} \;\;;
%\label{Z}
%\end{equation}
and  the probability to find a cell with a volume $v$ reads
\begin{equation}
P(v) = \frac{1}{V/T - v^*} \left( 1 - \frac{v-v^*}{V - Tv^*}\right)^{T-1} \;\;.
\label{Pv0}
\end{equation}
In the (thermodynamic) limit $T \rightarrow \infty$, this expression simplifies to
\begin{equation}
P(v) = \frac{1}{ \left< v \right>  - v^{*} }  \exp( - \frac{v-v^*}{\left<   v  \right>  - v^*} ) \;\;;
\label{Pv1}
\end{equation}
with $\left<v \right>=V/T$ the average volume.

\begin{figure}
\begin{center}
\resizebox{.4\textwidth}{!}{%
\includegraphics{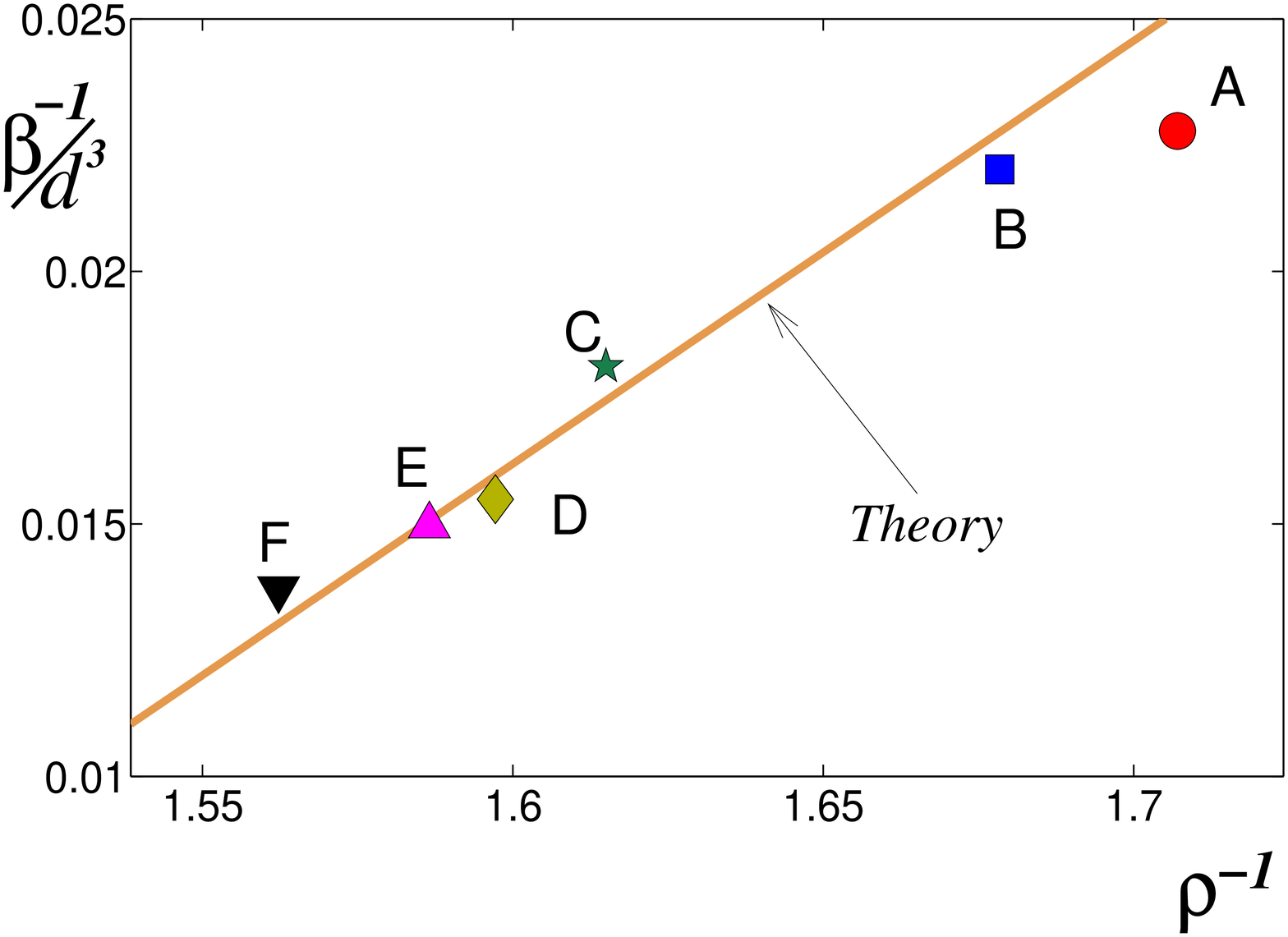}
}
\end{center}
\caption{ 
Coefficents $\beta^{-1}/d^3$ v.s. the inverse packing fraction $\rho^{-1}$.
The full symbols correspond to the six samples A-F.
The line is the  theoretical predictions from Eq.\ref{betaA}.}
\label{f.betarho}
\end{figure}

The empirical analysis  of the six samples A-F, reveals that -indeed- the volumes of the Delaunay tetrahedra follow distributions which, at large volumes, are well described by the exponential behavior  $P(v)\propto \exp(-\beta v ) $ with the coefficients $\beta$ ranging between $\beta \simeq 44/d^3 $ at $\rho=0.586$ (sample A) to $\beta \simeq 73/d^3 $ at $\rho=0.64$ (sample F)  (see Figs.\ref{f.DeVol} and \ref{f.betarho}).
Equation \ref{Pv1} predicts: 
\begin{equation}
\beta^{-1} =  { \left< v \right> - {v^*}  } = \frac{2}{\left< f \right>-2}\frac{V}{N} - v^* \;\;\;.
\label{beta}
\end{equation}
The expected values of $\beta$ can be obtained by imposing the three criteria on  mechanical stability, geometrical constraints and space filling. 
In particular, the geometrical constraint gives $v^* =   \sqrt{2}d^3/12$.
The criterium for mechanical stability imposes $\left< f \right> \simeq 14.5$.
The space filling condition implies $\rho = \pi d^3 N/(6 V)$.
By substituting these values, Eq. \ref{beta} gives
\begin{equation}
\beta^{-1}  \simeq \frac{\pi d^3 }{37.5}  \rho^{-1}  -  \frac{ \sqrt{2} d^3}{12}\;\;\;.
\label{betaA}
\end{equation}
A comparison between the prediction from this expression and the empirical results is shown in Fig.\ref{f.betarho} where $\beta^{-1}/d^3 $ is plotted versus $\rho^{-1}$.
The agreement between the theory and the experimental data is remarkable especially if one considers that there are no adjustable parameters.
Note that other choices for the elementary volumes, such as  the Vorono\"{\i} decomposition, do not  yield to exponential behaviors in the volume distributions \cite{Aste05rev}.

%\vspace*{3cm}
\begin{figure}
\begin{center}
\resizebox{.4\textwidth}{!}{%
\includegraphics{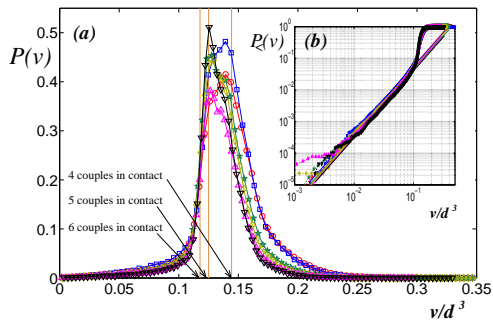}
}
\end{center}
\caption{ 
(a) Distribution of Delaunay volumes.
The vertical lines indicate the maximum volumes attainable by tetrahedra with 4, 5, or 6 couples of spheres in contact.
They are respectively: $\sqrt{3}d^3/12$, $d^3/8$ and $\sqrt{2}d^3/12$.
(b) Log-log plots of the cumulant distributions.
}
\label{f.DeVol1}
\end{figure}

Probably the most striking property of granular materials is their eclectic behavior which is nether classifiable as a solid nor as a fluid \cite{Jaeger92}.
It has been shown perviously that the structure of such systems can be conveniently described in term of a packing of tetrahedra with probability distribution of large volumes which follow an exponential behavior.
On the other hand, a more detailed analysis of the volume distribution, shown in Fig.\ref{f.DeVol1}, reveals that the exponential  behaviour is followed at large volumes only; whereas a more complex shape characterizes small volumes. 
From Figs.\ref{f.DeVol} and \ref{f.DeVol1}  one identifies that the volumes at which the distribution chases to be exponential are in a range between $\sim 0.12d^3$ and $\sim 0.14d^3$. 
Meaningfully, this range of volumes corresponds to Delaunay tetrahedra where most of the couples of spheres are in contact.
In particular, when all the 6 couples spheres are in touch the volume is $v^*\simeq 0.118d^3$;  
whereas, when 5 couples are in touch a tetrahedron can reach a maximum volume of  $d^3/8 = 0.125d^3$; conversely, when only 4 spheres are in touch the maximum reachable volume is $\sqrt{3}d^3/12 \simeq 0.144d^3$.
This fact indicates that below a given volume the tertraheda are made of spheres in contact and geometrical constraints become unavoidable and relevant. 
One can therefore argue that  these systems can be conveniently viewed as comprised of two phases:  
1) a phase made by compact tetrahedra ($v  <  0.144d^3$) which are geometrically constrained and are responsible for the mechanical stability; 
2) a phase made by loose tetrahedra ($v  >  0.144d^3$) which are geometrically unconstrained and take volumes accordingly with the distribution in Eq.\ref{Pv1}. 

Let me note  that some tetrahedra can assume very small volumes, even down to $10^-3d^3$ (see Fig.\ref{f.DeVol1}). 
A Delaunay tetrahedron with zero volume corresponds to a configuration of 4 in-plane spheres.
Therefore, configurations with volumes smaller than $v^*$ are, in general, loose packings.
Remarkably, Fig.\ref{f.DeVol1}b reveals that the probability distribution for such small volumes follows a power law beahviour with typical exponents between 1.09 and 1.17.
Such power laws, might be related with the power laws observed in the distributions of the radial distance between couples of spheres \cite{AstePRE05}.

In conclusion, by means of two independent methods (namely the analysis of the dihedral angles and the study of the volume distribution),  I have shown that sphere packs can be conveniently studied  as space-filling assemblies of elementary tetrahedra.
I have demonstrated that the volumes of such tetrahedra follow  an exponential distribution (at large volumes) which is controlled by the three conditions of mechanical stability, geometrical constraints and space-filling.
It has been discussed that the system's state can be described in term of the coefficient at the exponent $\beta$ which is analogous to Edwards' compactivity $(\lambda X)^{-1}$.
The theoretical predictions for  $\beta$  is in very good agreement with the empirical observations.
Such an agreement is particularly remarkable considering that there are no adjustable parameters.
The analysis of the probability distribution at small volumes reveals that, below $v \simeq 0.144 d^3$,  geometrical constraints, associated to the non-overlapping condition, lead to a more complex distribution which is shaped by the accessible configurations in systems of touching spheres.
Such differences in the kind of distributions at large and small volumes, is a signature of structural heterogeneity.
The granular system exploits such heterogeneity maximizing entropy and freedom while constraining mechanical stability.

Many thanks to T.J.  Senden and M. Saadatfar  for the tomographic data and several discussions.
This work was partially supported by the ARC discovery project DP0450292.

\end{document}